\documentclass[aps, reprint, superscriptaddress,
 amsmath,amssymb,
 aps,
prl,
floatfix,
]{revtex4-1}

\usepackage{graphicx}
\usepackage{dcolumn}
\usepackage{bm}
\usepackage{xfrac}
\usepackage{txfonts}
\usepackage{siunitx}
\usepackage{upgreek}
\usepackage[version=4]{mhchem}
\usepackage{float}
\usepackage{titlesec}
\titlespacing*{\section}{0pt}{1.1\baselineskip}{\baselineskip}
\usepackage{hyperref} 
\hypersetup{
colorlinks=true,
citecolor=black,
linkcolor=black,
filecolor=black,
urlcolor=black,
}
\usepackage{comment}
\begin{document}
\title{Strong Aharonov-Bohm quantum interference in simply-connected \ce{LaAlO3}/\ce{SrTiO3} structures}

\author{Patrick Irvin}
\affiliation{Department of Physics and Astronomy, University of Pittsburgh, Pittsburgh, PA 15260, USA}
\author{Hyungwoo Lee}
\author{Jung-Woo Lee}
\affiliation{Department of Materials Science and Engineering, University of Wisconsin-Madison, Madison, Wisconsin 53706, USA}
\author{Megan Briggeman}
\author{Shicheng Liu}
\author{Anil Annadi}
\author{Guanglei Cheng}
\author{Michelle Tomczyk}
\author{Jinanan Li}
\author{Mengchen Huang}
\affiliation{Department of Physics and Astronomy, University of Pittsburgh, Pittsburgh, PA 15260, USA}
\author{Chang-Beom Eom}
\affiliation{Department of Materials Science and Engineering, University of Wisconsin-Madison, Madison, Wisconsin 53706, USA}
\author{Jeremy Levy}
\email{jlevy@pitt.edu}
\affiliation{Department of Physics and Astronomy, University of Pittsburgh, Pittsburgh, PA 15260, USA}

\date{\today}

\begin{abstract}
We report Aharonov-Bohm (AB)-type quantum interference in simply-connected devices created at the \ce{LaAlO3}/\ce{SrTiO3} interface using conductive-atomic force microscope (c-AFM) lithography. The oscillations are multi-periodic functions of magnetic field strength, and they exhibit a substantial magnetic hysteresis with frequencies that depends on the magnetic sweep direction. The oscillation amplitude for the lowest two frequencies approaches $e^2/h$, consistent with the theoretical maximum for the AB effect, and harmonics up to third order are observable. Broadband quasiperiodic behavior is reported in a fraction of simply-connected electron waveguide devices that exhibit magnetic asymmetries. Curiously, nanoscale ring devices that are multiply-connected lack signatures of AB quantum interference. The interference phenomena are associated with an inhomogeneous magnetic landscape within the \ce{LaAlO3}/\ce{SrTiO3} nanostructures.

\end{abstract}

\maketitle

The Aharonov-Bohm (AB) effect is a quantum-mechanical interference phenomenon in which electrons couple directly to the electromagnetic vector potential \cite{Aharonov1959,Batelaan2009}. The AB effect is characterized by conductance oscillations that are periodic in the magnetic flux passing through the loop ($\Phi_0 = B \cdot A = h/e$, where $B$ is the magnetic field and $A$ is the area of the loop). AB interference contributes to a wide range of mesoscopic quantum transport phenomena \cite{Heinzel2007}, including universal conductance fluctuations \cite{Lee1985a} and weak localization/anti-localization. AB-related effects are distinct from other types of quantum oscillations, e.g., Shubnikov-de Haas, which is a precursor to the quantum Hall effect and is not a periodic function of magnetic field or flux. AB interference oscillations were first reported in the solid state by Webb et al. \cite{Webb1985a} and have since been reported for III-V nanostructures \cite{Timp1987}, graphene rings \cite{Russo2008}, and cross sections of carbon nanotubes \cite{Bachtold1999}. The maximum amplitude of conductance oscillations is bounded by the conductance quantum $G_0 = e^2/h$, but generally is an order of magnitude, or more, smaller. 

The complex-oxide heterostructure \ce{LaAlO3}/\ce{SrTiO3} \cite{Ohtomo2004,Sulpizio2014,Pai2018a} supports a two-dimensional conductive interface with an unusually large cohort of strongly-coupled phases, including superconductivity \cite{Reyren2007,Caviglia2008}, magnetism \cite{Brinkman2007,Pai2018a,Ariando2011,Dikin2011,Li2011a,Bert2011}, and ferroelasticity \cite{Honig2013}. A variety of quantum interference effects have been observed at oxide interfaces, in particular the \ce{LaAlO3}/\ce{SrTiO3} system. Weak localization and antilocalization have been reported, as well as universal conductance fluctuations in mesoscopic devices \cite{Rakhmilevitch2010}. Critical current oscillations were reported in an \ce{LaAlO3}/\ce{SrTiO3} ring device, signifying superconducting quantum interference device (SQUID) behavior \cite{Goswami2016}. The mobility of the 2D \ce{LaAlO3}/\ce{SrTiO3} interface is generally quite modest compared with other well-investigated systems like modulation-doped GaAs/AlGaAs \cite{Timp1987b} or graphene \cite{Russo2008}. It is therefore surprising that quasi-1D channels formed at the \ce{LaAlO3}/\ce{SrTiO3} interface exhibit highly ballistic transport: Tomczyk et al. have created 1D cavities that exhibit Fabry-Perot quantum interference \cite{Tomczyk2016a}, and recent reports by Annadi et al. show evidence for ballistic mean-free paths of order $l_{MF} \sim 20\,\upmu\textnormal{m}$ \cite{Annadi2018}. 

\begin{figure}
    \centering
    \includegraphics{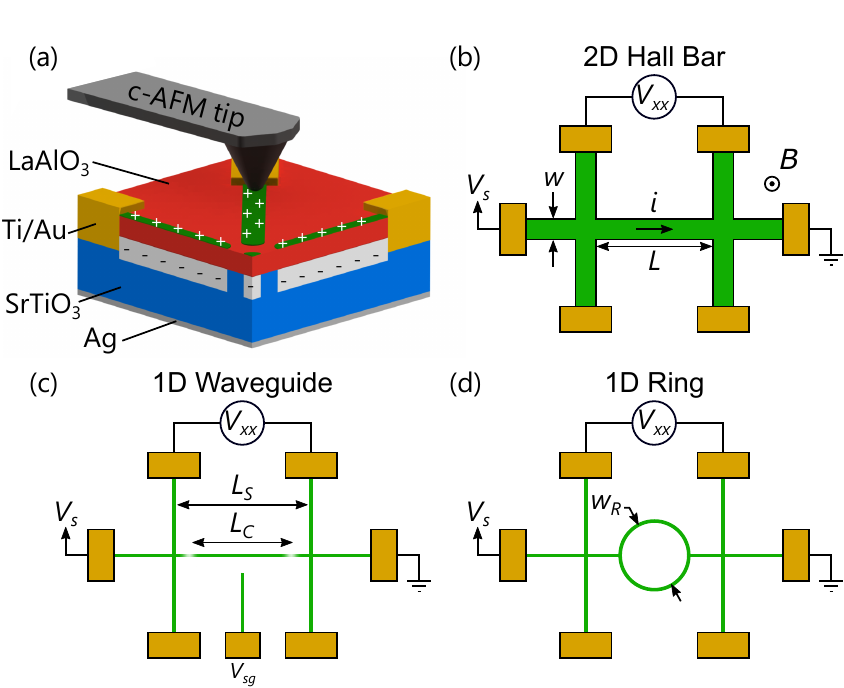}
    \caption{Experimental diagram. (a) Schematic of c-AFM lithography technique. (b) 2D Hall Bar and transport measurement scheme used for all device geometries. $V_s$: source bias; $V_{xx}$: longitudinal voltage; $i$: current; $B$: out-of-plane magnetic field; $w$: width of channel; $L$: length of channel; (c) 1D Waveguide. $L_S$: spacing between barriers; $L_C$: length of main channel (d) 1D Ring. $w_R$: diameter of ring.}
    \label{fig:fig1}
\end{figure}

Here we investigate low-temperature magnetotransport in three classes of \ce{LaAlO3}/\ce{SrTiO3} nanodevices: (i) 2D Hall bars, (ii) quasi-1D electron waveguides, and (iii) quasi-1D rings. All of the devices are created from \ce{LaAlO3}/\ce{SrTiO3} heterostructures with 3.4 unit cells of \ce{LaAlO3} grown by pulsed laser deposition on TiO$_2$-terminated \ce{SrTiO3} (100) substrates. 
Details of growth and processing are described elsewhere \cite{Pai2018a}. Conductive nanostructures are created using conductive atomic-force microscope (c-AFM) lithography (Figure \ref{fig:fig1}(a)). In this technique, a conductive AFM tip is scanned on the \ce{LaAlO3} surface, depositing positively-charged ions (mostly \ce{H+}) \cite{Bi2010,Brown2016} that locally switch the \ce{LaAlO3}/\ce{SrTiO3} interface to a conductive phase. The insulating phase can be restored locally by changing the sign of the applied voltage to the AFM tip. 
Conductive nanostructures are cooled in a dilution refrigerator and magnetotransport experiments are performed near its base temperature, $T \sim 50$ \si{mK}. The three device geometries investigated are illustrated in Figure \ref{fig:fig1}(b-d). 

\begin{figure*}
    \centering
    \includegraphics{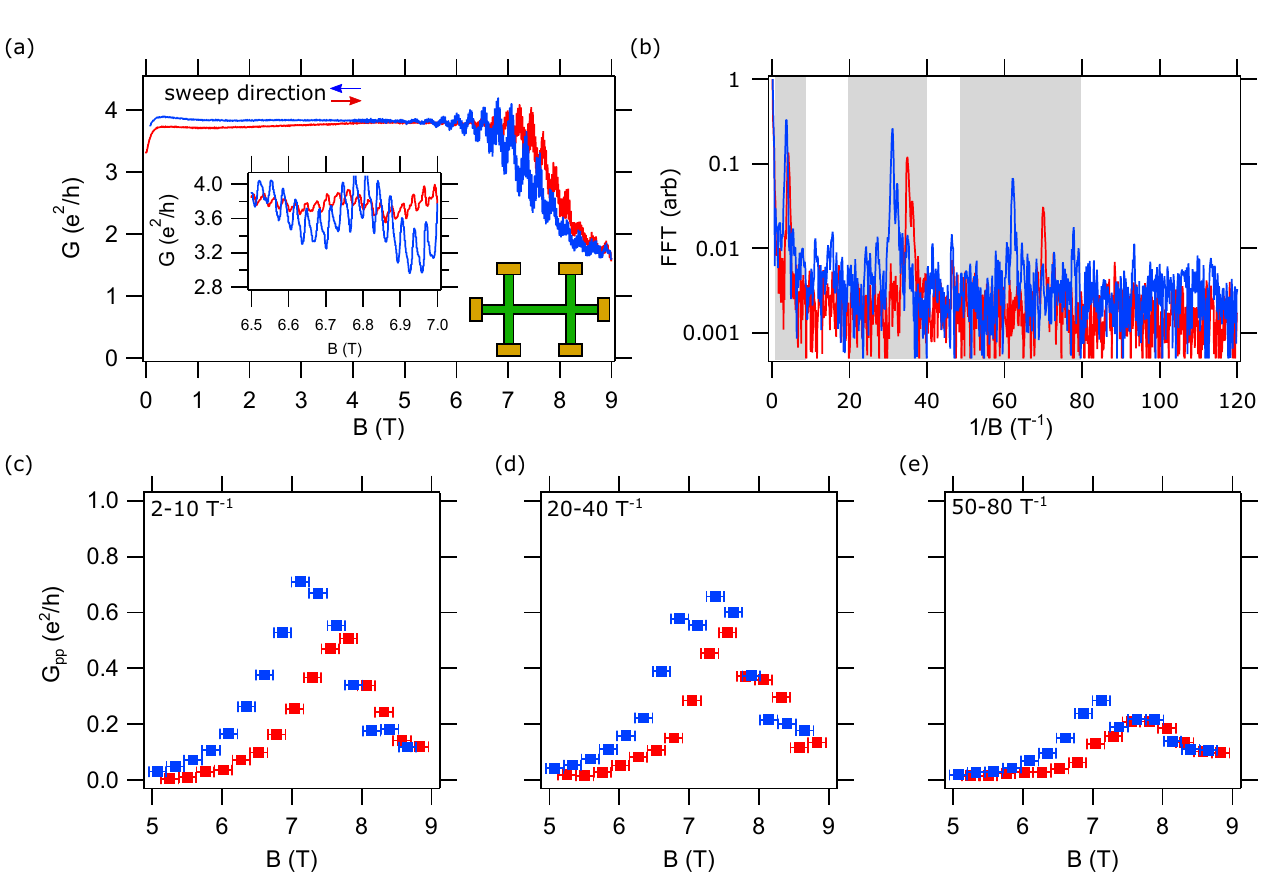}
    \caption{2D Hall Bar. (a) Magnetoconductance $G$ shown as a function of increasing (decreasing) magnetic field $B$, indicated by red (blue) curves. Sweep rate is 10 Oe/s. Inset shows multiperiodic oscillations in the magnetic field range 6.5 - 7 T. (b) Corresponding power spectra for increasing (decreasing) field sweeps. (c)-(e) The peak-peak amplitude of conductance ($G_{pp}$) after performing a bandpass filter in the ranges indicated by shaded regions shown in panel (b): 2-10 \si{\per\tesla} (c), 20-40 \si{\per\tesla} (d), and 50-80 \si{\per\tesla} (e). $A = 0.375 \,\upmu\textnormal{m}^2$ leads to an expected period of 90 \si{\per\tesla}.}
    \label{fig:fig2}
\end{figure*}

We first consider a Hall bar device, illustrated in Figure \ref{fig:fig1}(b). Key dimensions are the channel width $w = 250$ \si{nm} and length $L = 1.5 \, \upmu \textnormal{m}$. Figure \ref{fig:fig2}(a) shows the four-terminal conductance measured as a function of the out-of-plane magnetic field $B$. The red trace indicates a magnetic field sweep ($dB/dt = 0.001 \textup{ T/sec}$) from 0 to 9 T, while the blue trace indicates the reverse sweep back to 0 T.
For the positive field sweep, the conductance is constant at $\sim 4 \, e^2/h$ until around 5 T, where oscillations appear that are periodic with respect to B and grow steadily in amplitude. These oscillations reach a maximum amplitude around 7 \si{\tesla} and subsequently begin to decrease. The overall conductance decreases to approximately $1.7 \, e^2/h$ at the maximum field of 9 \si{\tesla}. For the negative field sweep, the low-frequency oscillations initially begin in-phase, but begin to deviate at around 8 \si{\tesla}. The conductance increases as the magnetic field strength decreases, and again there are multi-periodic oscillations. However, the average conductance trace shows approximately 1 \si{\tesla}-wide hysteresis. As will be discussed below, the frequency of oscillations for the negative field sweep is systematically lower than for the upward sweep. A comparison of the magnetic field sweep in the range 6.5 to 7 \si{\tesla} (Figure \ref{fig:fig2}(a), inset) illustrates how the two curves fail to align with one another. At fields below 5 \si{\tesla}, there are overall ($\sim 5\%$) changes in the conductance, as can be seen from differences between the forward and reverse traces in the range 1 to 4 \si{\tesla}.

A Fourier transform of the conductance oscillations (Figure \ref{fig:fig2}(b)) reveals several distinct frequencies. The largest amplitude (over the range 6 to 8 \si{\tesla}) is close to $e^2/h$, with a forward (increasing $B$) frequency of $f_1^+ = 4.3$ \si{\per\tesla}, and a reverse (decreasing $B$) frequency $f_1^- = 3.7$ \si{\per\tesla}. There is a higher frequency $f_2^+ = 34$ \si{\per\tesla} in the forward direction, reducing to $f_2^- = 31$ \si{\per\tesla} in the reverse direction. These peaks show strong second harmonic at $2f_2^+$ and $2f_2^-$, respectively, and a faint third harmonic at $3f_2^+$. There are other smaller peaks which only show up for one peak direction and are significantly smaller (e.g., at 46 \si{\per\tesla} and 78 \si{\per\tesla}).

In order to quantify the amplitude of oscillations over small $B$ field ranges, we also calculated the peak-to-peak conductance, $G_{pp}$, shown in Figure \ref{fig:fig2}(c-e). The magnetoconductance is first bandpass-filtered with upper and lower bounds indicated by the shaded regions in Figure \ref{fig:fig2}(b). The peak-to-peak amplitude is then calculated for the filtered traces. The bandpass filter ranges for panels (c), (d), and (e) are 2-10 \si{\per\tesla}, 20-40 \si{\per\tesla}, and 50-80 \si{\per\tesla}, respectively. It is striking to note that the maximum oscillation amplitude is more than $0.7 \,e^2/h$ for the reverse $B$ sweep direction near 7 T.

\begin{figure}
    \centering
    \includegraphics{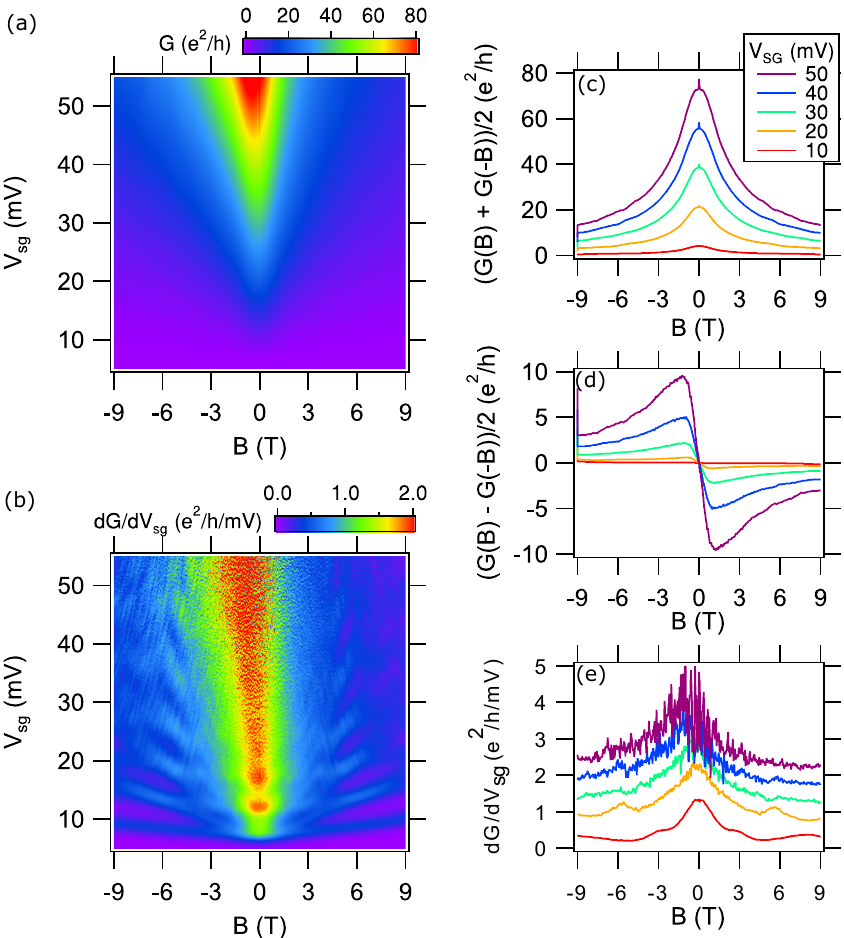}
    \caption{1D Electron waveguide. (a) Conductance $G$ vs magnetic field $B$ and side gate $V_{sg}$. (b) Transconductance $dG/dV_{sg}$ showing interference effects. (c) Symmetric and (d) antisymmetric components of s of conductance for various values of $V_{sg}$. (e) Transconductance of various values of $V_{sg}$. $T = 50 \,\textnormal{mK}$.}
    \label{fig:fig3}
\end{figure}

The next category of device is an electron waveguide. This class of device has been investigated elsewhere \cite{Annadi2018} and shows highly ballistic quantized transport, associated with subbands that are characterized by well-defined vertical, lateral, and spin degrees of freedom. The subband bottoms are marked by quantized changes in conductance, by peaks in the transconductance $dG/dV_{sg}$, and can also vary with magnetic field. Most electron waveguides exhibit transport characteristics with no evidence of AB interference. However, a small fraction ($\sim10\%$ of devices) exhibit a gate-dependent magnetic-field asymmetric magneto-conductance and associated gate-dependent oscillations with magnetic field. Figure \ref{fig:fig3}(a) shows an intensity plot of conductance of one such device (main channel length $L_C = 1800 \,\textnormal{nm}$, barrier spacing $L_S = 1000 \,\textnormal{nm}$, main channel width $w \sim 10 \,\textnormal{nm}$) as a function of sidegate voltage $V_{sg}$ and out-of-plane magnetic field $B$, at $T = 50 ~ \textnormal{mK}$. With increasing $V_{sg}$, the conductance transitions from being highly symmetric with $B$ to developing a significant asymmetric contribution. The symmetrized and antisymmetrized responses are shown in Figures \ref{fig:fig3}(c) and (d). The first two curves show negligible asymmetry, but above $V_{sg} \sim 20 \,\textnormal{mV}$, the asymmetry grows. Figure \ref{fig:fig3}(c) shows evidence of superconductivty near zero magnetic field and high $V_sg$. Rapid oscillations with magnetic field are apparent in the corresponding transconductance map $dG/dV_{sg}$ (Figure \ref{fig:fig3}(e)). The transconductance shows, in addition to the signatures of emergent subbands with increasing $V_{sg}$, pronounced oscillations that vary both with $B$ (at fixed $V_{sg}$) and with $V_{sg}$ (at fixed $B$). Figure \ref{fig:fig3}(b) illustrates the emergence of conductance oscillations and their asymmetry with magnetic field sign. The magnitude of these oscillations is approximately a factor of two larger for $B < 0$ compared with $B > 0$. The conductance fluctuations are not periodic, but they are reproducible for multiple $B$ field sweeps on a given device, and they persist at temperatures where the electron waveguide subband structure is no longer visible. Figure S1 
\cite{Supplement} shows data for the same waveguide at $T = 600 \,\textnormal{mK}$: the subband features are hardly visible, and yet the gate-dependent asymmetry and conductance fluctuations remain.

The third class of device is a “ring” structure (Figure \ref{fig:fig2}(d)), similar to the original design of Webb et al. \cite{Webb1985a}. The width of the ring $w_R$ is 500 nm, corresponding to an expected frequency of 47 \si{\per\tesla} (assuming a fundamental flux periodicity $h/e$). Figure S2
(a) shows magnetoconductance at different back gate voltages, and Figure S2
(b) shows the corresponding Fourier spectrum for one such device. (Data shown is representative of more than 20 multiply-connected devices tested.) There is no evidence for AB quantum interference oscillations as the magnetic field is varied. More than a dozen similar devices have been investigated, and none of them exhibit any quantum interference. Also, none of the devices have exhibited magnetic asymmetries such as those shown in Figure \ref{fig:fig2}.

Here we summarize and discuss the main results. Two categories of simply-connected nanostructures (2D Hall Bars and 1D Waveguides) exhibit rapid oscillations, which appear to be associated with AB-derived interference. That is, the oscillations are periodic in $B$ and not $1/B$, as would be the case for Shubinkov-de Haas oscillations. Cheng et al. have argued elsewhere \cite{Cheng2018} that the Shubnikov-de Haas oscillations reported in \ce{LaAlO3}/\ce{SrTiO3} by other groups \cite{Caviglia2010d,BenShalom2010} are in fact explainable in terms of magnetic depopulation of magnetoelectric subbands. The data we present is consistent with this interpretation. Both structures exhibit either magnetic hysteresis or significant gate-tunable magnetic asymmetries. The multiply-connected ring structure shows no signature of AB oscillations. One remarkable aspect of the AB interference in the Hall bar structure is the magnitude of the oscillations. The peak-to-peak amplitude of the $f_1$ oscillation reaches nearly $0.8 \, e^2/h$, which is close to the theoretical maximum. The higher frequency $f_2$ shows a magnitude that is nearly as large. In addition, there is a sizeable second harmonic ($h/2e$ periodic) response and even a faint third-harmonic response ($h/3e$ periodic). Specifically, we find a correlation between the AB oscillation amplitude and the average value of the conductance (which is comparable to the conductance quantum.) We can infer the phase coherence of the paths that would lead to AB interference by examining the amplitude of the AB interference amplitude ($\sim 0.3 \, e^2/h$) and estimating the minimum perimeter that can give rise to an interference effect with area $A = 1.41 \,\upmu\textnormal{m}^2$. A straightforward analysis yields a phase coherence length of $0.7 \,\upmu\textnormal{m}$, which is significantly larger than what one expects from a 2D mobility of $\sim 10^3  \,\textnormal{cm}^2\textnormal{/Vs}$.

The fact that AB interference is observed in an open Hall bar structure indicates that the current flow within is both highly ballistic and highly filamentary. The four-terminal conductance (in the 2-4 $e^2/h$ range) provides a quantitative estimate of the number of active channels. The fact that the AB oscillation magnitude is as large (of order $e^2/h$) as the conductance is decreasing indicates that there is an inhomogeneous landscape and there is an energy-dependent “beamsplitter” for exactly one mode of propagation for the electrons. Both paths of the interferometer must be highly ballistic for interference to be so clean. Below 105 K there is a ferroelastic symmetry breaking in \ce{SrTiO3} that leads to extended line defects at the \ce{LaAlO3}/\ce{SrTiO3} interface \cite{Honig2013,Kalisky2013}. Frenkel et al \cite{Frenkel2017} have shown that current can travel preferentially along these naturally occurring ferroelastic domain boundaries. A random configuration of domain boundaries within the Hall bar may lead to AB interference being observed in some devices but not others. We also note that the c-AFM lithography process can seed $z$-oriented ferroelastic domains, which also could lead to filamentary conduction, however the precise domain configuration is unknown and will be the subject of future studies.

We note a qualitative similarity with simulations of AB oscillations for a graphene-based ring device by Dauber et al. \cite{Dauber2017}. Two requirements are needed to exhibit interference: (i) the chemical potential and magnetic field need to be tuned to a region where there is magnetic depopulation, and (ii) there needs to be an inhomogeneous region at either the inlet or the outlet of the device to serve as the “beam splitter”. Dauber observes two distinct AB frequencies: a low frequency corresponding to orbital motion around the inhomogeneous region, and a higher frequency corresponding to the larger ring area. Following this interpretation (and setting aside all of the other obvious differences between the two material systems), we surmise that the low frequency represents the area defined by the beamsplitting region, while the higher frequency corresponds to the area that defines the two paths within the Hall bar channel. The rise and fall of both frequencies together suggest that the transport is correlated. The existence of magnetism in this device (but not in most others) indicates that there are magnetic “patches”, similar to those that have been imaged by scanning SQUID microscopy \cite{Bert2011}. The intersection of a magnetic patch with the Hall bar would be consistent with both the hysteretic magnetoresistance and the difference in AB frequencies for positive and negative sweep directions, as observed in metallic rings by Sekiguchi et al \cite{Sekiguchi2008}. The existence of the local magnetization could strongly spin split the electron subbands, possibly leading to a quantum anomalous Hall regime \cite{Annadi2018,Chang2013a} with associated ballistic transport and long-range interference effects.

While the electron waveguide device is significantly narrower than the Hall bar device, magnetic asymmetries are again correlated with the interference effects. The interference is stable at much higher temperatures than those for which the subband structure is visible, again indicating that the magnetic landscape is dominating over the waveguide itself. Here, instead of having a single or small number of distinct frequencies, the magnetic response is consistent with universal conductance fluctuations. The “bandwidth” ($\Delta T^{-1}$) of these fluctuations is bounded by the area of the electron waveguide, including all of the available transverse modes of the waveguide. Many such devices have been investigated elsewhere \cite{Annadi2018}, and most do not exhibit universal conductance fluctuations. The only devices that show UCF also exhibit some kind of magnetic anisotropy, again indicating that magnetic scattering plays an important role in creating multiple paths for quantum interference effects.

Finally, we discuss the absence of quantum interference in the ring device. The device itself does not show evidence of magnetic hysteresis or asymmetry. More than twenty other devices with multiply-connected geometries have been investigated, none of which exhibit signatures of AB interference. However, it is still surprising, given that AB interference can be observed in this material system, that a multiply-connected device would not exhibit any interference effects. There is one significant difference between the two devices that show AB interference and the ring structure: both are simply connected devices. That is to say, the region separating the two paths for the electron is conducting. Therefore, the electrons themselves are subject to short-range screening, preventing the electron travelling along one path from “seeing” or sensing the other path. The ring geometry consists of two paths that are separated by a highly polarizable \ce{SrTiO3} medium. The Thomas-Fermi screening length in \ce{LaAlO3}/\ce{SrTiO3} can be large in the insulating \ce{SrTiO3}.  However, within the conducting \ce{SrTiO3} itself, the screening should be highly effective.  If we consider the electron traveling along one of the two paths, there will be a sizeable dielectric response that could provide “which-path” information, and associated decoherence.

\begin{acknowledgments}
This work is supported by the U.S. Department of Energy, Office of Basic Energy Sciences under Award Number \MakeUppercase{DE-SC}0014417. J.L. acknowledges support from the Vannevar Bush Faculty Fellowship program sponsored by the Basic Research Office of the Assistant Secretary of Defense for Research and Engineering and funded by the Office of Naval Research (N00014-15-1-2847). The work at the University of Wisconsin-Madison was supported by AFOSR under Grant No. FA9550-15-1-0334.
\end{acknowledgments}


%

\end{document}